\documentclass[12pt]{article}
\usepackage{epsfig}
\def\beq{\begin{equation}}
\def\eeq{\end{equation}}
\def\beqa{\begin{eqnarray}}
\def\eeqa{\end{eqnarray}}
 
\newlength{\dinwidth} \newlength{\dinmargin}
\begin{document}

\begin{center}
{\Large \bf Two-loop soft anomalous dimensions and NNLL resummation for heavy quark production}
\end{center}
\vspace{2mm}
\begin{center}
{\large Nikolaos Kidonakis}\\
\vspace{2mm}
{\it Kennesaw State University, Kennesaw, GA 30144, USA}
\end{center}
 
\begin{abstract}
I present results for two-loop soft anomalous dimensions for heavy quark 
production which control soft-gluon resummation at 
next-to-next-to-leading-logarithm (NNLL) accuracy. 
I derive an explicit expression for the exact result and study it 
numerically for top quark production via $e^+ e^- \rightarrow t {\bar t}$, 
and I construct a surprisingly simple but very accurate approximation. 
I show that the two-loop soft anomalous 
dimensions with massive quarks display a simple proportionality relation 
to the one-loop result only in the limit of vanishing quark mass.
I also discuss the extension of the calculation to single top and top pair
production in hadron colliders.
\end{abstract}
 
\thispagestyle{empty} \newpage \setcounter{page}{2}
 
The calculation of higher-order corrections to heavy quark cross sections and differential 
distributions is needed for increasing the accuracy of the theoretical predictions. 
Given the importance of the top quark \cite{top} to electroweak and Higgs physics, the top quark 
cross section at the Tevatron and the LHC, both in top pair \cite{ttbar}
and single-top \cite{singletop} production modes, 
as well as at a future $e^+ e^-$ collider needs to be calculated with the highest possible 
precision.
Bottom and charm quark cross sections are also important in 
understanding QCD and calculating backgrounds to new physics. 

Soft-gluon emission is an important contributor to higher-order corrections, particularly 
near partonic threshold. The soft-gluon corrections can be formally resummed to all orders in 
perturbation theory. The resummation arises from the factorization of the cross section into a 
hard-scattering function $H$ and a soft function $S$ that describes noncollinear soft-gluon 
emission in the process \cite{NKGS}. 
The evolution of the soft function  is controlled by a process-dependent soft anomalous dimension 
$\Gamma_S$. The resummed cross section can be written in moment space as 
$\sigma^{\rm res}(N)=\exp[E_{i}(N,\mu)] H(\mu) S(N) 
\exp[\int_m^{m/N} (d\mu/\mu) \,  2\, \Gamma_S(\mu)]$, 
with $\mu$ the factorization scale, $m$ the heavy quark mass, and 
$N$ the moment variable which is conjugate to a kinematical variable 
measuring distance from partonic threshold. 
The exponents $E_i$ resum collinear and soft 
radiation from any massless partons in the scattering. The logarithms 
of $N$ exponentiate. At leading-logarithm accuracy  we determine the 
coefficients of the highest power of $\ln N$ at each order in the strong 
coupling $\alpha_s$, 
at next-to-leading logarithm (NLL) accuracy we also determine the 
second highest power of $\ln N$, etc. 

For heavy quark pair, such as $t {\bar t}$, hadroproduction the soft anomalous dimensions  
are matrices in color space due to the complicated color structure of the scattering 
processes, and were calculated at one loop for both $q{\bar q}\rightarrow t{\bar t}$ and 
$gg\rightarrow t{\bar t}$ channels in 
Ref. \cite{NKGS}, thus allowing resummation at next-to-leading logarithm.
From the calculation in \cite{NKGS} one can also derive the corresponding result for 
heavy quark production at $e^+ e^-$ machines, a simple function related to the cusp 
anomalous dimension of Ref. \cite{KorRad}. 
For single top production at hadron colliders the one-loop soft anomalous dimensions 
were presented for the $t$, $s$, and $tW$ channels
in Ref. \cite{NKsingletop}. 
To increase the accuracy of the resummation to next-to-next-to-leading logarithms 
(NNLL) for all these processes one needs to derive the soft anomalous dimensions at 
two loops. 

The calculations of soft anomalous dimensions involve diagrams 
with eikonal (Wilson) lines representing the heavy quarks.
The eikonal approximation is valid for describing the emission
of soft gluons from particles in the hard scattering 
and leads to a modified form of the Feynman rules and thus of diagram 
calculations (see e.g. [4-9]).
When the gluon momentum goes to zero, the quark-gluon
vertex reduces to $g_s T_F^c \, v^{\mu} / v\cdot k$, 
with $g_s$ the strong coupling, $v$ a dimensionless velocity vector, 
$k$ the gluon momentum,
and $T_F^c$ the generators of SU(3) in the fundamental representation.
Below we calculate explicitly the soft anomalous dimension through two loops
for the process $e^+ e^- \rightarrow t {\bar t}$, study its properties in 
detail and then discuss the extension to processes with more complicated 
color structure, i.e. single top and top pair hadroproduction. 

\begin{figure}
\begin{center}
\includegraphics[width=15cm]{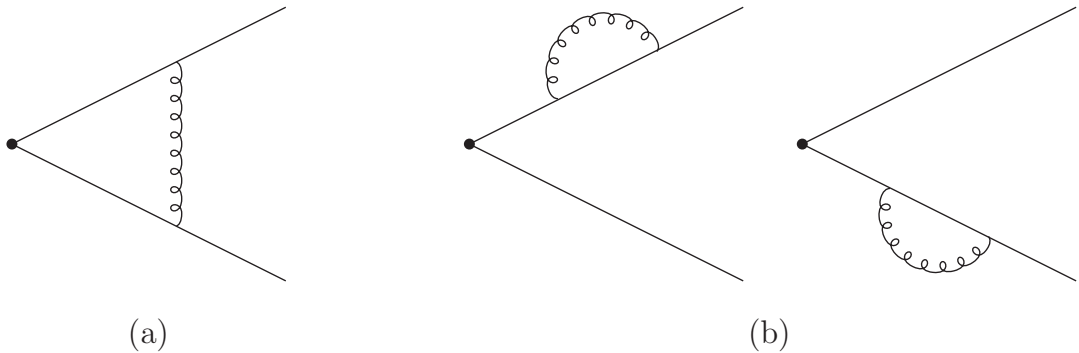}
\caption{One-loop diagrams with heavy-quark eikonal lines.}
\end{center}
\label{1loop}
\end{figure}

The eikonal diagrams are calculated in dimensional regularization with 
$n=4-\epsilon$ and in Feynman gauge in momentum 
space. The one-loop soft anomalous dimension, $\Gamma_S^{(1)}$, can be read 
off the coefficient of the ultraviolet (UV) pole of the one-loop diagrams 
in Fig. 1. Writing 
$\Gamma_S=(\alpha_s/\pi) \Gamma_S^{(1)}+(\alpha_s/\pi)^2 \Gamma_S^{(2)}+\cdots$, 
we find the one-loop expression 
\beq
\Gamma_S^{(1)}=C_F \left[-\frac{(1+\beta^2)}{2\beta} 
\ln\left(\frac{1-\beta}{1+\beta}\right) -1\right]
\label{GammaS1}
\eeq
with $C_F=4/3$, $\beta=\sqrt{1-4m^2/s}$, and 
$s=(p_{e^+}+p_{e^-})^2$.

\begin{figure}
\begin{center}
\includegraphics[width=15cm]{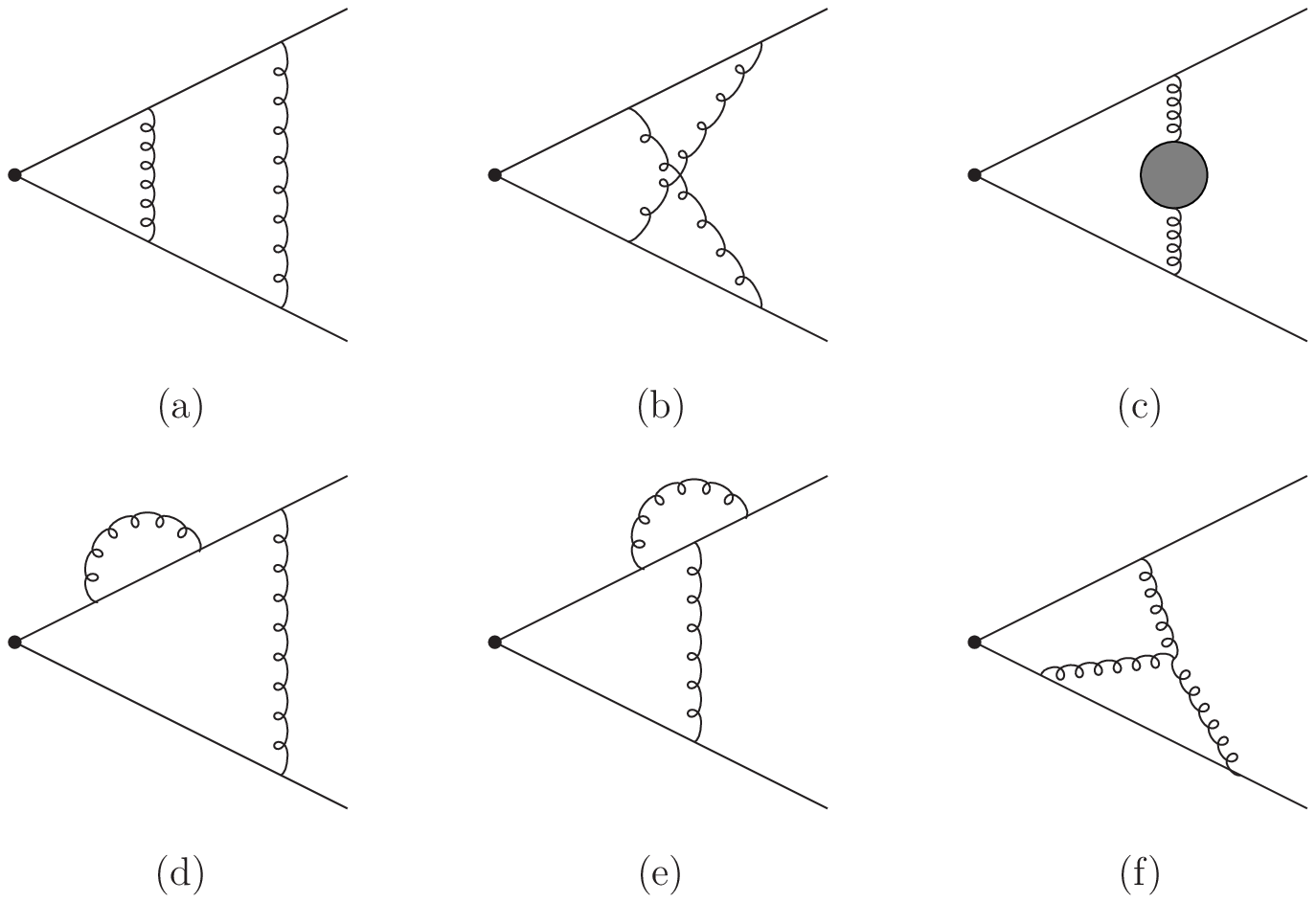}
\caption{Two-loop vertex diagrams with heavy-quark eikonal lines.}
\end{center}
\label{2vloop}
\end{figure}

\begin{figure}
\begin{center}
\includegraphics[width=15cm]{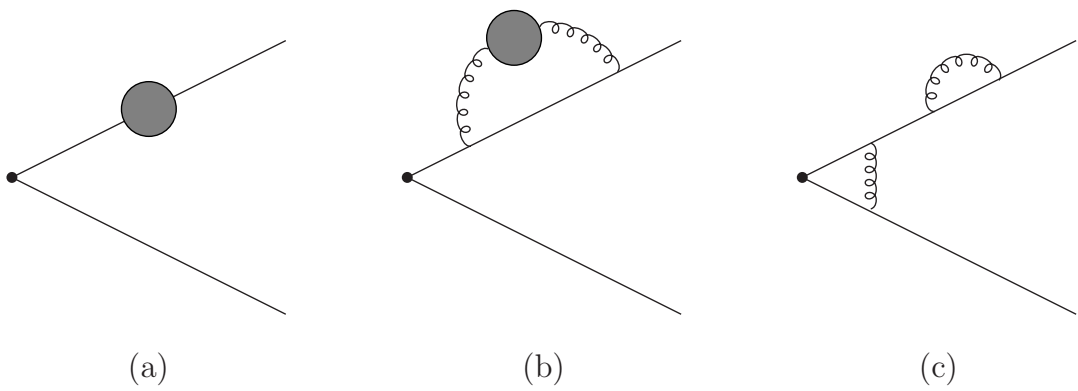}
\caption{Two-loop heavy-quark self-energy diagrams with eikonal lines.}
\end{center}
\label{2sloop}
\end{figure}

We now continue with the two-loop diagrams.  
A few results for some of these diagrams have appeared 
in \cite{NK2l,NKPS,NKPANIC08}.
In Fig. 2 we show graphs with vertex 
corrections and in Fig. 3 graphs with heavy-quark self-energy corrections. 
There are three more graphs, not shown,
with vanishing contributions: two graphs with gluon loops involving 
a four-gluon vertex, and one graph involving a three-gluon vertex with  
all three gluons attaching to a single eikonal line. The calculations are 
challenging due to the heavy quark mass and involve multiple complicated  
integrals and delicate separations 
of infrared and ultraviolet poles (which by construction of the 
soft function are opposites of each other \cite{NKGS}). 
The analytical structure involves 
logarithms and polylogarithms. It is understood in the results below 
that for each diagram we 
include the appropriate one-loop counterterms for the divergent subdiagrams
(but note that graph 2(b) does not have a divergent subdiagram). 
Typically there are large cancellations of terms between the 
individual diagrams and their respective counterterms.

We now present the UV poles of the kinematic terms for the two-loop 
diagrams, which later will have to be combined with color and symmetry 
factors, and an overall factor $\alpha_s^2/\pi^2$.
The sum of the UV poles of diagrams 2(a) and 2(b) is minus one-half the square of the UV pole of the one-loop 
diagram of Fig. 1(a): 
$I_{2a}+I_{2b}=[(1+\beta^2)^2/(8 \beta^2)] (-1/\epsilon^2) \ln^2[(1-\beta)/(1+\beta)]$.
The UV poles of the crossed diagram, Fig. 2(b), are
\beq
I_{2b}=\frac{(1+\beta^2)^2}{8\beta^2} 
\frac{1}{\epsilon} \left\{-\ln\left(\frac{1-\beta}{1+\beta}\right)
\left[{\rm Li}_2\left(\frac{(1-\beta)^2}{(1+\beta)^2}\right)
+\zeta_2\right] 
-\frac{1}{3}\ln^3\left(\frac{1-\beta}{1+\beta}\right)
+{\rm Li}_3\left(\frac{(1-\beta)^2}{(1+\beta)^2}\right)-\zeta_3 \right\}.
\nonumber \\
\eeq

The diagram in Fig. 2(c) represents corrections from quark, 
gluon, and ghost loops. The quark-loop contribution is
$I_{2cq}=[n_f (1+\beta^2)/(6\beta)]
[1/\epsilon^2-5/(6 \epsilon)] \ln[(1-\beta)/(1+\beta)]$
with $n_f$ the number of light quark flavors, while the gluon plus ghost contribution is 
$I_{2cg}=(5/24) [(1+\beta^2)/\beta]  
[1/\epsilon^2-31/(30 \epsilon)] \ln[(1-\beta)/(1+\beta)]$.
Diagram 2(d) is given by
\beqa
I_{2d}&=&\frac{(1+\beta^2)}{4\beta}
\left\{-\frac{1}{\epsilon^2} \ln\left(\frac{1-\beta}{1+\beta}\right)
+\frac{1}{\epsilon} \left[\frac{1}{2}\ln^2\left(\frac{1-\beta}{1+\beta}\right)
+\ln\left(\frac{1-\beta}{1+\beta}\right)
\right. \right.
\nonumber \\ &&  \left. \left.
{}+\ln\left(\frac{1-\beta}{1+\beta}\right) 
\ln\left(\frac{(1+\beta)^2}{4\beta}\right)
-\frac{1}{2}{\rm Li}_2\left(\frac{(1-\beta)^2}{(1+\beta)^2}\right)
+\frac{\zeta_2}{2}\right] \right\} 
\eeqa
and the result for diagram 2(e) is its negative: $I_{2e}=-I_{2d}$.
The diagram of Fig. 2(f) gives 
\beqa
I_{2f}=\frac{1}{\epsilon} \left\{-\frac{1}{4} \left[2 \zeta_2+\ln^2\left(\frac{1-\beta}{1+\beta}\right)\right] 
\left[\frac{(1+\beta^2)}{2 \beta} \ln\left(\frac{1-\beta}{1+\beta}\right)+1\right]
+\frac{(1+\beta^2)}{12 \beta} \ln^3\left(\frac{1-\beta}{1+\beta}\right) \right\} \, .
\eeqa

Diagram 3(a) represents two graphs 
with overall $C_F^2$ color factor, 
$I_{3a1}=-3/(2 \epsilon^2)+1/(2\epsilon)$,
and another graph 
$I_{3a2}=1/\epsilon^2-1/(2 \epsilon)$.
The quark-loop contribution to graph 3(b) is
$I_{3bq}=(n_f/3) [1/\epsilon^2-5/(6 \epsilon)]$
while the gluon plus ghost loop contribution is 
$I_{3bg}=(5/12)[1/\epsilon^2-31/(30 \epsilon)]$.
Note the relation $d/d\beta [2\beta I_{2c}/(1+\beta^2)]
=[-2/(1-\beta^2)] I_{3b}$.
Finally diagram 3(c) gives
$I_{3c}= [(1+\beta^2)/(2\beta)] (-1/\epsilon^2) 
\ln[(1-\beta)/(1+\beta)]$.

Combining the above kinematic results with color and symmetry factors, 
the contribution of the diagrams in Figs. (2) and (3) to the two-loop soft 
anomalous dimension is  
\beqa
&&
C_F^2 \left[I_{2a}+I_{2b}+2 \, I_{2d}
+2 \, I_{2e}+I_{3a1}+I_{3a2}+I_{3c}\right]
\nonumber \\ 
{}&+&C_F \, C_A \left[-\frac{1}{2} I_{2b} 
+I_{2f} -I_{2cg}
-I_{2e}-I_{3bg}-\frac{1}{2} I_{3a2} \right]
+\frac{1}{2} C_F \left[I_{2cq}+I_{3bq}\right]  
\nonumber \\
&=&
-\frac{1}{2 \epsilon^2} \left(\Gamma_S^{(1)}\right)^2
+\frac{\beta_0}{4 \epsilon^2} \Gamma_S^{(1)}
-\frac{1}{2 \epsilon} \Gamma_S^{(2)} 
\label{S2}
\eeqa
where we have cancelled out factors of $\alpha_s^2/\pi^2$ on both 
sides of the equation.
On the right-hand side of Eq. (\ref{S2})  
in addition to $\Gamma_S^{(2)}$, which appears in the coefficient of the  
$1/\epsilon$ pole, there also  appear terms from the exponentiation 
of the one-loop result and the running of the coupling, 
with $\beta_0=(11/3) C_A-2n_f/3$, 
$C_A=3$,  which account for all the double poles of the graphs.
From Eq. (\ref{S2}) we solve for the two-loop soft anomalous dimension:
\beqa
\Gamma_S^{(2)}&=&\left\{\frac{K}{2}+\frac{C_A}{2} 
\left[-\frac{1}{3}\ln^2\left(\frac{1-\beta}{1+\beta}\right)+\ln\left(\frac{1-\beta}{1+\beta}\right)-\zeta_2\right] \right.
\nonumber \\ && \hspace{15mm} \left.
+\frac{(1+\beta^2)}{4 \beta} C_A \left[{\rm Li}_2\left(\frac{(1-\beta)^2}
{(1+\beta)^2}\right)+\frac{1}{3}\ln^2\left(\frac{1-\beta}{1+\beta}\right)+\zeta_2\right]\right\} \, \Gamma_S^{(1)}
\nonumber \\ && \hspace{-18mm}
{}+C_F C_A \left\{\frac{1}{2}
+\frac{1}{2} \ln\left(\frac{1-\beta}{1+\beta}\right)
+\frac{1}{3} \ln^2\left(\frac{1-\beta}{1+\beta}\right)
-\frac{(1+\beta^2)^2}{8 \beta^2} \left[
-{\rm Li}_3\left(\frac{(1-\beta)^2}{(1+\beta)^2}\right)+\zeta_3\right] \right.
\nonumber \\ && \left.
{}-\frac{(1+\beta^2)}{2 \beta} \left[\ln\left(\frac{1-\beta}{1+\beta}\right) 
\ln\left(\frac{(1+\beta)^2}{4 \beta}\right)-\frac{1}{6}\ln^2\left(\frac{1-\beta}{1+\beta}\right) 
-{\rm Li}_2\left(\frac{(1-\beta)^2}{(1+\beta)^2}\right)\right]\right\}.
\label{Gammas2}
\eeqa
We have written the two-loop result $\Gamma_S^{(2)}$ 
in Eq. (\ref{Gammas2}) in the form of a term which is a multiple 
of the one-loop soft anomalous dimension $\Gamma_S^{(1)}$ 
plus additional terms.
The well-known two-loop constant $K$ \cite{JKLT} is given by 
$K=C_A (67/18-\zeta_2)-5n_f/9$.
Note that the color structure of $\Gamma_S^{(2)}$ 
involves only the factors $C_F C_A$ and $C_F n_f$; all $C_F^2$ 
terms cancel out. 

\begin{figure}
\begin{center}
\includegraphics[width=12cm]{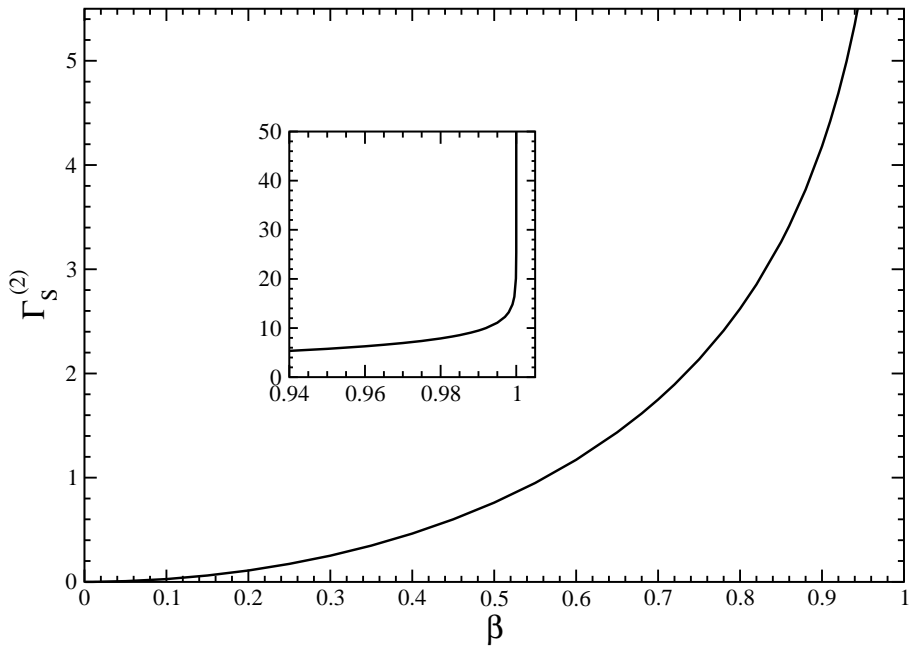}
\caption{Two-loop soft anomalous dimension $\Gamma_S^{(2)}$ for top quark production via
$e^+ e^-\rightarrow t{\bar t}$.}
\label{Gamma2}
\end{center}
\end{figure}

The two-loop soft anomalous dimension for $e^+ e^- \rightarrow t {\bar t}$ 
is plotted as a function of 
$\beta$ with $n_f=5$ in Fig. 4. Note that $\Gamma_S^{(2)}$ vanishes at 
$\beta=0$, the threshold limit, and diverges at $\beta=1$, 
the massless limit. 
The inset in Fig. 4 shows more clearly the sharp increase of $\Gamma_S^{(2)}$
as $\beta$ approaches 1. Figure 5 also displays $\Gamma_S^{(2)}$ but in a 
logarithmic plot which makes the small $\beta$ region more clear.

\begin{figure}
\begin{center}
\includegraphics[width=12cm]{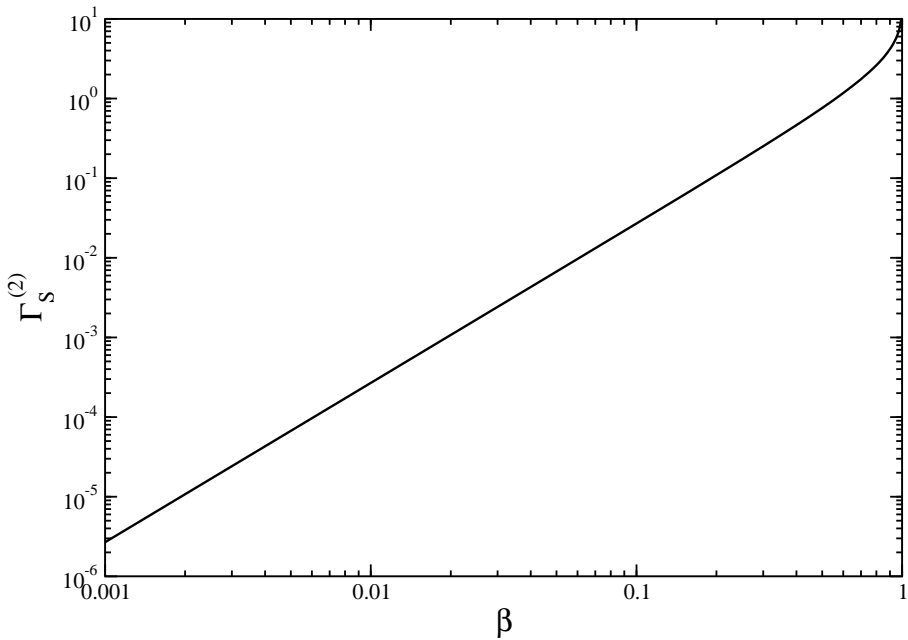}
\caption{Logarithmic plot of $\Gamma_S^{(2)}$ for top quark production via 
$e^+ e^- \rightarrow t{\bar t}$.}
\label{Gamma2log}
\end{center}
\end{figure}

It is instructive to study the small and large $\beta$ behavior of $\Gamma_S^{(2)}$ 
in more detail. We first expand around $\beta=0$ and find 
\beqa
\Gamma_{S \, {\rm exp}}^{(2)}&=&-\frac{2}{27} \beta^2 
\left[C_F C_A (18 \zeta_2-47)+5 C_F n_f\right]+{\cal O}(\beta^4) \, .
\label{Gammasexp}
\eeqa
Note that $\Gamma_S^{(2)}$ is an even function of $\beta$ and hence 
only even powers of $\beta$ appear in the expansion. In Fig. 6 we 
see that just keeping the first term in the expansion as shown in Eq. (\ref{Gammasexp})  
provides a good approximation to the complete result for small $\beta$.
The ratio of the expansion to the complete $\Gamma_S^{(2)}$ remains within a few 
percent of unity for values of $\beta$ up to around 0.3.
Keeping more powers of $\beta$ in the expansion results of course in 
a better approximation. Keeping powers through $\beta^{12}$ results in 
an excellent approximation to $\Gamma_S^{(2)}$ over a wide range of $\beta$,
starting to differ by a few percent above $\beta=0.8$ as seen from the 
corresponding curve in Fig. 6.   

As observed in Ref. \cite{ADS}, the two-loop soft anomalous dimension for 
processes with massless quarks is proportional to the one-loop result 
of Ref. \cite{KOS} by the 
simple factor $K/2$, i.e. $\Gamma_{S\, {\rm massless}}^{(2)}
=(K/2) \, \Gamma_{S\, {\rm massless}}^{(1)}$. 
This property was further explored and generalized in Refs. \cite{BN,GM,LD}.
We see this $K/2$ factor for heavy quarks as the first term in 
Eq. (\ref{Gammas2}).
It is clear that for the massive case the situation is rather more 
complicated than for massless processes: although all the $C_F n_f$ terms 
of $\Gamma_S^{(2)}$  are included in $(K/2) \, \Gamma_S^{(1)}$, there are 
many additional $C_F C_A$ terms.
It is interesting to check however that in the massless limit of our 
$\Gamma_S^{(2)}$ this relation 
is satisfied. In Fig. 6 we also plot the ratio of $(K/2) \, \Gamma_S^{(1)}$ to 
$\Gamma_S^{(2)}$. We see that indeed the ratio is 1 at the limit $\beta=1$, 
i.e. the massless limit, so the massless proportionality relation is recovered.
However, for any other value of $\beta$ the ratio is not one; in fact it is 
1.144 near $\beta=0$, the threshold limit, gradually decreases with increasing
$\beta$, and falls steeply to 1 in the region very close to $\beta=1$. 

Given that the expansion around $\beta$, Eq. (\ref{Gammasexp}), gives very good  
approximations to $\Gamma_S^{(2)}$ at 
smaller $\beta$ while the expression $(K/2)\, \Gamma_S^{(1)}$ is a better 
approximation at large $\beta$, we can derive an approximation 
to $\Gamma_S^{(2)}$ for all $\beta$ values by starting with the $\beta$ 
expansion of $\Gamma_S^{(2)}$,  
and then adding $(K/2) \, \Gamma_S^{(1)}$ and subtracting from it its $\beta$ 
expansion: 
\beqa
\Gamma^{(2)}_{S \, {\rm approx}}&=&\Gamma^{(2)}_{S \, {\rm exp}}
+\frac{K}{2} \Gamma_S^{(1)}-\frac{K}{2} \Gamma^{(1)}_{S \, {\rm exp}}
\nonumber \\ &=&
\frac{K}{2} \Gamma_S^{(1)}+C_F C_A \left(1-\frac{2}{3}\zeta_2\right) \beta^2
+{\cal O}\left(\beta^4\right) \, .
\label{Gapprox}
\eeqa
As seen from Fig. 6, adding just the $\beta^2$ terms to 
$(K/2)\, \Gamma_S^{(1)}$ as in Eq. (\ref{Gapprox}) 
provides an excellent approximation to the exact result for $\Gamma_S^{(2)}$ 
for all $\beta$, never more than half a percent away from the exact value. 
It is quite remarkable that such a simple expression provides this 
good an approximation to a rather long and complicated expression 
for $\Gamma_S^{(2)}$.
If we keep additional terms through order $\beta^{12}$ then the approximation 
is truly impressive, not differing by more than one per mille anywhere in the 
$\beta$ range, and in fact not differing by more than a few parts per million for 
most $\beta$.  
At a linear collider with $\sqrt{s}=500$ GeV, $\beta=0.726$ for a top mass 
of $172$ GeV. For this value of $\beta$ the expansion through order 
$\beta^{12}$ of Eq. (\ref{Gammasexp}) as 
well as the approximate expression at order $\beta^2$ or higher 
of Eq. (\ref{Gapprox}) give excellent approximations. However, 
$(K/2) \Gamma_S^{(1)}$ by itself is 11\% larger than $\Gamma_S^{(2)}$.

\begin{figure}
\begin{center}
\includegraphics[width=12cm]{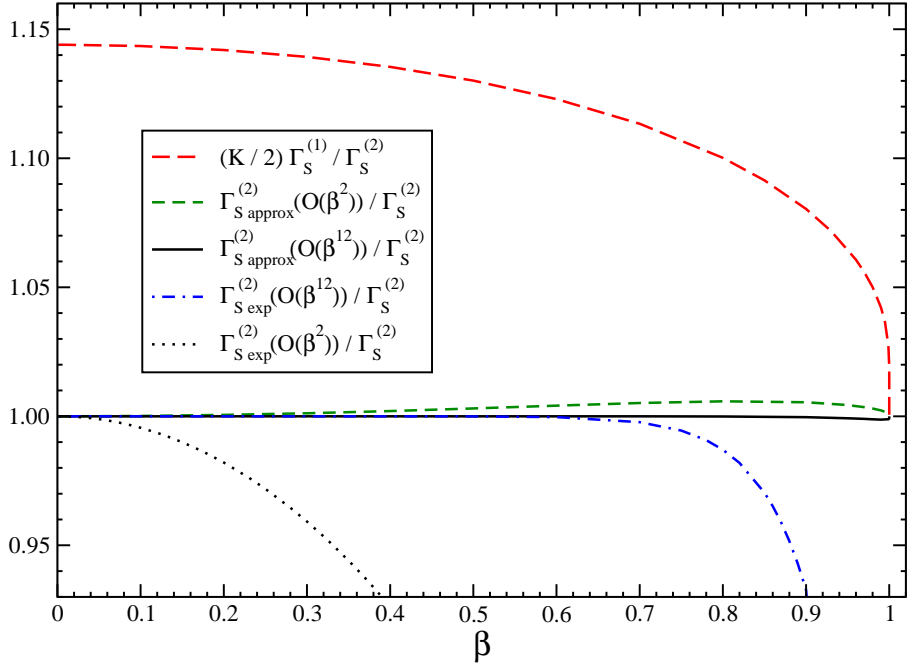}
\caption{Expansions and approximations to $\Gamma_S^{(2)}$ for 
$e^+ e^- \rightarrow t{\bar t}$.}
\label{Gamma2approx}
\end{center}
\end{figure}

The result in Eq. (\ref{Gammas2}) can be written in terms 
of the cusp angle $\gamma=\ln[(1+\beta)/(1-\beta)]$ as
\beqa
\Gamma_S^{(2)}&=&\frac{K}{2} \, \Gamma_S^{(1)}
+C_F C_A \left\{\frac{1}{2}+\frac{\zeta_2}{2}+\frac{\gamma^2}{2}
-\frac{1}{2}\coth^2\gamma\left[\zeta_3-\zeta_2\gamma-\frac{\gamma^3}{3}
-\gamma \, {\rm Li}_2\left(e^{-2\gamma}\right)
-{\rm Li}_3\left(e^{-2\gamma}\right)\right] \right.
\nonumber \\ && \hspace{25mm} \left.
{}-\frac{1}{2} \coth\gamma\left[\zeta_2+\zeta_2\gamma+\gamma^2
+\frac{\gamma^3}{3}+2\, \gamma \, \ln\left(1-e^{-2\gamma}\right)
-{\rm Li}_2\left(e^{-2\gamma}\right)\right] \right\},
\label{2lcusp}
\eeqa
where $\Gamma_S^{(1)}=C_F (\gamma \coth\gamma-1)$, and is consistent with 
the cusp anomalous dimension of Ref. \cite{KorRad}. 

Finally we consider the two-loop soft anomalous dimensions for single top and for 
top pair (or bottom or charm pair) production at hadron colliders.  
Since the two-loop soft anomalous 
dimensions for massless quarks obey the simple proportionality relation to one loop, 
and the diagrams for single top and top pair processes involve eikonal graphs with both
massless and massive eikonal lines,  
it is clear that for $\beta$ near 1 the relation 
$\Gamma_S^{(2)}=(K/2)\, \Gamma_S^{(1)}$ 
will hold for both single top and $t{\bar t}$ production. 
For smaller $\beta$, however, this relation is no longer valid. 
For reference, at the Tevatron and the LHC the $t{\bar t}$ cross section 
receives most contributions in the region around $0.3 < \beta < 0.8$ 
which peak roughly around $\beta \sim 0.6$. 
However, since even a 15\% discrepancy between the exact $\Gamma_S^{(2)}$ and 
$(K/2) \,\Gamma_S^{(1)}$ may result in less than 1\% effect in the total 
cross section it may turn out that using the massless proportionality 
relation is adequate for many purposes (see also the related 
discussion in \cite{NKRV}). 
Nevertheless, a rigorous treatment 
requires that we include the extra terms for 
the massive case. More details will be given elsewhere. 

This work was supported by the National Science Foundation under 
Grant No. PHY 0555372.

\end{document}